\documentclass[]{spie} 

\usepackage{amsmath}
\usepackage{wrapfig}
\usepackage{amsmath,amsfonts,amssymb}
\usepackage{graphicx}
\usepackage[colorlinks=true, allcolors=blue]{hyperref}

\title{Photonic chip for visible interferometry: laboratory characterization and comparison with the theoretical model}

\author[a]{Manon Lallement}
\author[a]{Sylvestre Lacour}
\author[a]{Elsa Huby}
\author[b]{Guillermo Martin}
\author[a]{Kevin Barjot}
\author[a]{Guy Perrin}
\author[a]{Daniel Rouan}
\author[a]{Vincent Lapeyrere}
\affil[a]{LESIA, Observatoire de Paris, Université PSL, CNRS, Sorbonne Université, Université Paris Cité, 5 place Jules Janssen, 92195 Meudon, France}

\author[c]{Sebastien Vievard}
\author[c,d]{Olivier Guyon}
\author[c]{Julien Lozi}
\author[c]{Vincent Deo}
\author[c]{Takayuki Kotani}
\affil[c]{The National Astronomical Observatory of Japan (NAOJ), 650 North A'ohoku Place, Hilo, Hawaii 96720, United States}
\affil[d]{University of Arizona, Tucson, AZ 85721, United States}

\affil[b]{University of Grenoble Alpes, CNRS, IPAG, 38000 Grenoble, France}

\author[e]{Cecil Pham}
\author[e]{Cedric Cassagnettes}
\author[e]{Adrien Billat}
\affil[e]{TEEM Photonics, 61 Chem. du Vieux Chêne, Meylan, France}

\author[f]{Nick Cvetojevic}
\affil[f]{Côte d'Azur Observatory, 96 Bd de l'Observatoire, 06300 Nice, France}

\author[g]{Franck Marchis}
\affil[g]{Search for Extra-Terrestrial Intelligence (SETI), 339 Bernardo Ave, Mountain View, CA 94043, United States}

\authorinfo{Further author information: \\Manon Lallement: E-mail: manon.lallement@obspm.fr, Telephone: +33 (0)6 23 98 44 56}

\begin{document} 
\maketitle

\begin{abstract}
Integrated optics are used to achieve astronomical interferometry inside robust and compact materials, improving the instruments stability and sensitivity. In order to perform differential phase measurements at the H$\alpha$ line (656.3nm) with the 600-800nm spectro-interferometer FIRST, a photonic integrated circuit (PIC) is being developed. This PIC performs the visible combination of the beams coming from the telescope pupil sub-apertures. In this work, TEEM Photonics waveguides fabricated by $K_+:Na_+$ ion exchange in glass are characterized in terms of single-mode range and mode field diameter. The waveguide diffused index profile is modeled on Beamprop software. FIRST beam combiner building blocks are simulated, especially achromatic directional couplers and passive $\pi/2$ phase shifters for a potential ABCD interferometric combination. 
\end{abstract}


\section{INTRODUCTION}
\label{sec:FIRST_Instrument}
\subsection{Pupil remapping and spectroscopy with FIRST}
FIRST, standing for Fibered Imager FoR a Single Telescope, was built by a team from Paris Observatory to validate the concept of pupil remapping \cite{perrin2006high}. This technique consists in recovering the source spatial intensity distribution with an angular resolution as fine as $\lambda/2D$, with  $\lambda/D$ the telescope diffraction limit and $D$ the diameter of the telescope primary mirror.

%To reach the diffraction limit by interferometry one can use the sparse aperture masking (SAM) technique. 
The sparse aperture masking (SAM) technique was proposed to restore the diffraction limit capability of a telescope thanks to interferometry. A mask with holes is put in the pupil plane. At the telescope focal plane, the image no longer corresponds to a classical point spread function %the well-known Airy diffraction pattern 
but to the superimposition of fringe patterns created by each pair of interfering sub-pupils beams. The non-redundant configuration of the mask means that each baseline, i.e each pair of sub-pupils separated by a baseline vector $ \vec{B} $, produces a unique fringe pattern. The information carried by each baseline can be retrieved independently. This is well illustrated in the Fourier domain, in which each baseline information, i.e phase and contrast of the associated fringes, is carried by a single peak isolated from the other. All this insures that there is no blurring effect between the fringes in presence of turbulence and that one can recover information at the diffraction limit of the telescope, and even below. A disadvantage of the SAM technique is that photons are lost due to the mask. The pupil remapping technique gives access to the whole pupil: the telescope pupil is divided into sub-pupils which are recombined in a way that the information at each baseline can be retrieved independently (non redundantly or pair-wise).

\begin{wrapfigure}{r}{9cm}
  \centering
  \includegraphics[height=9cm]{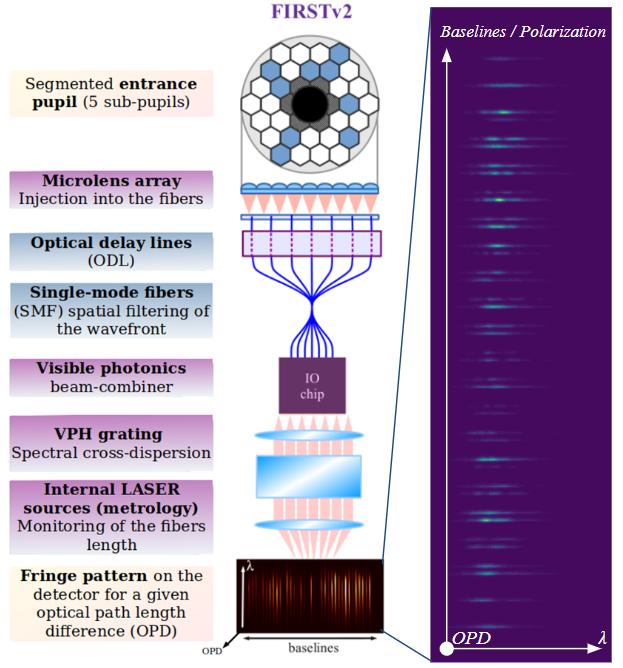}
  \caption[FIRSTv2SetUp] 
%>>>> use \label inside caption to get Fig. number with \ref{}
  { \label{fig:FIRSTv2SetUp} FIRSTv2 Set up}
\end{wrapfigure}

In the FIRST instrument, the telescope pupil is divided into sub-pupils thanks to a micro-lens array which couples the sub-pupil light into Single-Mode Fibers (SMF). SMF spatially filter the wavefront. Optical delay lines are used to compensate for the fiber length difference and reach the zero Optical Path Difference (OPD). 

In the previous version of FIRST, called FIRSTv1, SMF were used to remap the sub-pupils into a non-redundant configuration. In FIRSTv2, SMF inject the sub-pupil light into a photonic integrated circuit where the beams are recombined pair-wise. The chip outputs are spectrally dispersed thanks to a Volume Holographic Grating (VPH) based spectrograph. Its resolution is $R\sim 3600$ at $670$ $nm$, covering the $600$-$800$ $nm$ range. Orthogonal polarization are also separated thanks to a Wollaston prism to avoid fringe blurring. Fig.~\ref{fig:FIRSTv2SetUp} presents the data acquired on the camera: 5 sub-pupils are recombined, so 10 baseline are present. Each baseline appears four times, two times because of the 2x2 directional couplers used for the beam combination and another two times because both polarization are imaged. To further sample the fringes, they are temporally modulated (OPD axis in Fig.~\ref{fig:FIRSTv2SetUp}) applying piston command to a segmented mirror.

For each baseline, the phase of the fringes is measured to highlight asymmetries in the spatial intensity distribution of the source\cite{huby2012first}.

\subsection{Science at the Subaru telescope}
From 2010 to 2013, the instrument was installed on the 3m-Shane telescope at the Lick Observatory. Since 2013, it is often used on sky at the 8.2m Subaru telescope\cite{elsaphd}. FIRST is installed on the Subaru Coronagraphic Extreme Adaptive Optics platform (SCExAO)\cite{jovanovic2015,SPIE_2021-Vievard}. The stable wavefront allows long time exposures, up to 1000ms, and observation of faint targets, see Table~\ref{tab:SCExAO}. 

 \begin{table}[h!]
\caption{Capabilities on SCExAO} 
\label{tab:SCExAO}
\begin{center}       
\begin{tabular}{|l|l|}
\hline
\rule[-1ex]{0pt}{3.5ex} Strehl delivered by SCExAO & 50$\%$ to 60$\%$ at 750 nm \\
\hline
\rule[-1ex]{0pt}{3.5ex} Number of sub-pupils used & 2x9 (FIRSTv1) and 5 (FIRSTv2) \\
\hline
\rule[-1ex]{0pt}{3.5ex} Field of view & $\sim$ 100 mas \\
\hline
\rule[-1ex]{0pt}{3.5ex} Angular resolution $\lambda$/D & $\le$ 16.5mas at 656.3 nm\\
\hline                                      
\rule[-1ex]{0pt}{3.5ex} On-sky current magnitude limit & 6.6\\
\hline

\end{tabular}
\end{center}
\end{table}

Until now, FIRSTv1 has been detecting close binary stars\cite{huby2013first}. Now, the scientific objective is to push FIRST detection limits to resolve exoplanetary systems. Young gas giant exoplanets are particularly interesting for FIRSTv2. Fernandes et al.\cite{fernandes2019} demonstrated that the gas giant distribution is supposed to be maximal for star-exoplanets systems with a separation of 1-3 A.U., corresponding to an angular separation between 7 and 28 mas at 140 pc (the distance of the Taurus Nuclear Cloud). These angular separations correspond to the diffraction limit of 8m class telescopes in the visible and are reached by FIRST at the Subaru telescope. Moreover, Seager and Deming\cite{seager2010} stated that instruments must reach a $10^{6}$ to $10^{9}$ dynamic in the visible to differentiate the light reflected by an exoplanet from its host-star by direct imaging. This is currently out of reach for FIRST. However, at the protoplanet state ($\le$ $4 Myr$ old), gas giants are still accreting matter from the surrounding disk and emit signal at the H$\alpha$ line ($656.3$ $nm$)\cite{aoyama2018,aoyama2019}. The required dynamic in this particular line is lowered down to $10^{2}$ to $10^{3}$. Therefore, very young gas giants are potentially here and easier to detect in the visible. For now, only three detections have been confirmed thanks to H$\alpha$ imaging. It was the case of protoplanets PDS70b and PDS70c with the MUSE integral field spectrograph in 2019\cite{wagner2018,haffert2019} and with AB Aurigae b in 2022 detected with the VAMPIRES instrument installed on SCExAO\cite{currie2022}.

H$\alpha$ differential phase measurement is a method spotlighting this accretion signal with spectro-interferometry. It consists in comparing the phase of the fringes at the H$\alpha$ line (where the protoplanet is detected) with the phase of the continuum (where it is not detected). This technique has been recently implemented using the high precision phase measurements with the GRAVITY instrument to detect the broad line region around a quasar\cite{Gravity2018}. To perform this measurement, FIRST spectral resolution, sensitivity and dynamic have to be enhanced. 

\section{FIRSTv2 upgrade: TEEM photonics integrated circuit for high performance beam combination}
\label{sec:FIRST_PIC}
\subsection{FIRSTv2 beam combination}
\subsubsection{Why a PIC ?}
 In order to detect accreting protoplanets by measuring their H$\alpha$ signal with FIRST: 1) the instrument photon throughput needs to be increased up to about a magnitude limit of 12 mag in the R band (PDS70). To meet this sensitivity specification a high throughput PIC is needed. Its transmission must be greater than 50$\%$ or 75$\%$ depending on the spectral band. 2) FIRST dynamic must reached $10^{2}$ - $10^{3}$ at the H$\alpha$ line. The dynamic performance directly depends on the phase measurement precision. 
 
 To enhance the stability and the precision on interferometric observables and meet the specifications listed in Table.~\ref{tab:SpecContrastPhase}, a high throughput visible beam combiner is being developed and characterized\cite{martin2016,martin2018}. 

 \begin{table}[h!]
\caption{FIRST visible PIC specifications} 
\label{tab:SpecContrastPhase}
\begin{center}       
\begin{tabular}{|l|l|l|}
\hline
\rule[-1ex]{0pt}{3.5ex} Wavelength & 650 to 660 nm & 640-650nm and 660-780nm \\ 
\hline
\rule[-1ex]{0pt}{3.5ex} Global transmission & $\ge 75\%$ & $\ge 50\%$ \\ 
\rule[-1ex]{0pt}{3.5ex} Phase accuracy & 0.1$^\circ$ & No specification \\
\hline
\end{tabular}
\end{center}
\end{table}

Compared to beam combination with bulk optics, photonic chips combination is more robust, less sensitive to thermal variations, mechanical constraints and alignment errors. The flux corresponding to one baseline interference is condensed in a few pixels, thus reducing the read-out noise. Using a PIC also makes it easier to increase the number of input sub-apertures, by densifying or duplicating the chip.
%At the moment, temporal modulation of the phase thanks to the segmented mirror is needed to properly sample the fringes. This constraints the acquisition frequency to about 20Hz to ensure that the fringe phase is stable during the modulation sequence, but --> j'ai vu que tu en parles après 

%Nonetheless, the gain in sensitivity is not straightforward. It is difficult to compare FIRSTv1 and FIRSTv2 in terms of throughput because the interferometric combination is different. In FIRSTv1, a non-redundant configuration of the sub-pupils was needed as the interferometric combination was performed in free space following a Young slit experiment like method. The fringes were spatially modulated, so that the OPD scan around the null value was performed in one direction and the spectral dispersion in the other direction on the camera. In FIRSTv2, fringes are temporally sampled by sending piston commands to the segmented mirror meaning that various images must be taken to scan the fringes around the null OPD. One baseline is imaged on fewer pixels so that the readout noise is reduced. Nonetheless, the amount of photon received by the camera is strongly dependent on the PIC throughput performance and on the atmospheric turbulence constraining the modulation frequency. The sensitivity gain between FIRSTv1 and FIRSTv2 will be further investigated.

\subsubsection{Specifications of a 5T beam combiner and required building blocks}

\begin{figure} [h]
   \begin{center}
   \begin{tabular}{c} %% tabular useful for creating an array of images 
   \includegraphics[height=6cm]{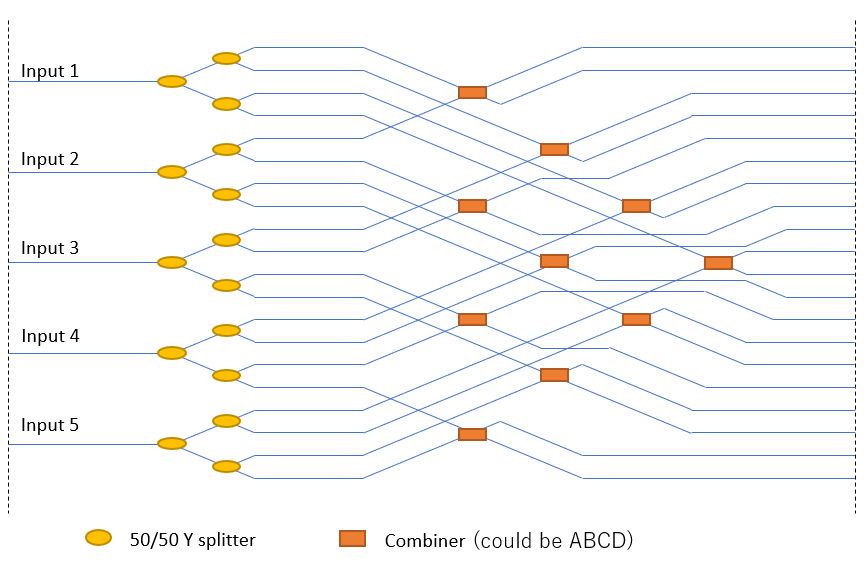}
   \end{tabular}
   \end{center}
   \caption[PICScheme] 
%>>>> use \label inside caption to get Fig. number with \ref{}
   { \label{fig:PICScheme} 
Interferometric beam combination scheme for FIRST 5T.}
   \end{figure}

The interferometric combination scheme for FIRST 5T consists in combining the light coming from 5 sub-pupils, pair by pair, as shown in  Fig.~\ref{fig:PICScheme}. Nufern PM-630-HP polarization maintaining fibers filter and inject the sub-pupils light into the chip. Each of the 5 inputs are split in four thanks to Y splitters and are then recombined with the other ones thanks to combiners. Combiners can be Y junctions (10 outputs, one per baseline), directional couplers (20 outputs) or ABCDs (40 outputs). 

In Table.~\ref{tab:SpecificationsPIC}, specifications on 50:50 achromatic splitting and coupling ratios are meant to maximize the fringes contrast over our spectral band. With directional couplers, a phase modulation is required to sample the fringes around the null OPD. This is currently done by sending piston commands to the segmented mirror following a 20 steps phase modulation sequence running at 20Hz. In an ABCD combination scheme, fringes are directly sampled on the four ABCD outputs as explained in Sec.~\ref{subsubsec:theoryABCD}. That way, phase modulation is shorter and acquisition time is reduced. 

 \begin{table}[h!]
\caption{Detailed FIRST visible PIC specifications. SW: straight waveguide. CW: crossed waveguide.} 
\label{tab:SpecificationsPIC}
\begin{center}       
\begin{tabular}{|l|l|l|l|}
\hline
\rule[-1ex]{0pt}{3.5ex} \textbf{Wavelength} & \textbf{640 to 670 nm }& \textbf{670 to 780 nm} \\
\hline
\rule[-1ex]{0pt}{3.5ex} \textbf{Monomodicity} & Yes & Yes\\
\hline
\rule[-1ex]{0pt}{3.5ex} \textbf{Total throughput} & $\ge$ 80$\%$ & $\ge$ 50$\%$\\
\hline                                      
\rule[-1ex]{0pt}{3.5ex} \textbf{Insertion loss }& $\le 0.13$ $dB$ & $\le 0.46$ $dB$ \\
\hline
\rule[-1ex]{0pt}{3.5ex} \textbf{Internal loss} & $\le 0.04$ $dB$ for SW and CW & $\le 0.04$ $dB$ for SW and $\le 0.13$ $dB$ CW\\
                                      & $\le 0.09$ $dB$ for Y junctions & $\le 0.27$ $dB$ for Y junctions \\
                                      & $\le 0.22$ $dB$ for directional couplers & $\le 0.7$ $dB$ directional couplers \\
\hline
\rule[-1ex]{0pt}{3.5ex} \textbf{Splitting ratio} & 50 $\pm$ 5$\%$ for Y junctions & 50 $\pm$ 15$\%$ for Y junctions\\
                                    & 50 $\pm$ 10$\%$ for directional couplers & 50 $\pm$ 30$\%$ for directional couplers\\
                                    & 25 $\pm$ $\%$ for ABCD & 25 $\pm$ $\%$ for ABCD\\
\hline                                      
\rule[-1ex]{0pt}{3.5ex} \textbf{Cross talk} & $\le$ 0.1$\%$ & $\le$ 0.1$\%$ \\     
\hline
\rule[-1ex]{0pt}{3.5ex} \textbf{Polarization maintaining} & Extinction $\le$ 1$\%$ & Extinction $\le$ 1$\%$\\     
\hline
\end{tabular}
\end{center}
\end{table}

\subsection{TEEM Photonics $K_+:Na_+$ ion exchange technology}
\subsubsection{Technology and first laboratory measurements}
\label{sec:Labmeas}
\begin{figure} [h!]
   \begin{center}
   \begin{tabular}{cc} %% tabular useful for creating an array of images 
   \includegraphics[height=6cm]{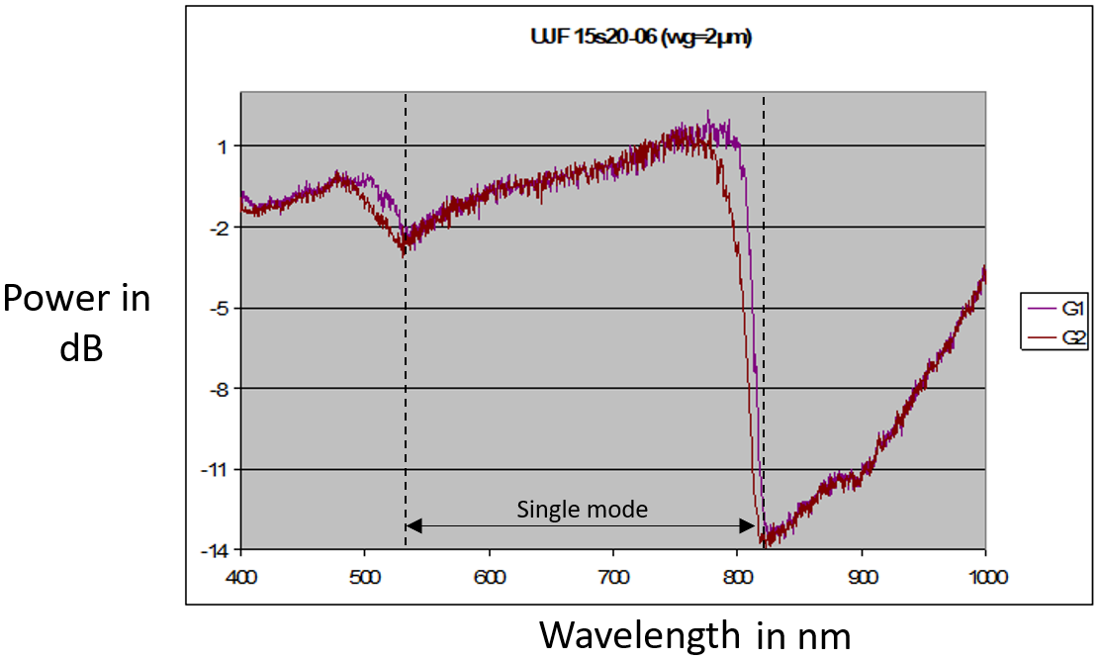}&\includegraphics[height=6cm]{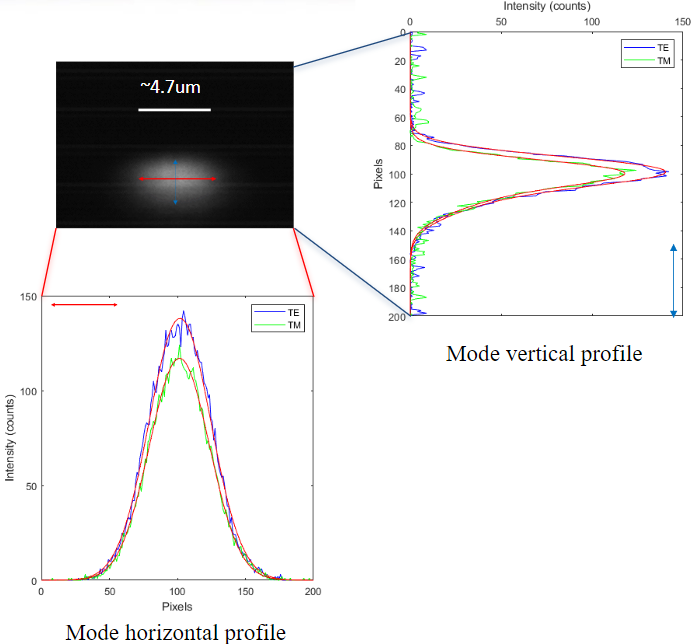}
   \end{tabular}
   \end{center}
   \caption[CutoffTEEM] 
%>>>> use \label inside caption to get Fig. number with \ref{}
   { \label{fig:cutoff} \textbf{Left:} Cutoff measurements for a straight waveguide built with a 2 $\mu m$ wide lithographic mask. \textbf{Right:} Mode field diameter measurements at $635$ $nm$. The straight waveguide output is imaged with a x40 objective on a Thorlabs CDD. The pixel size is 5.2 $\mu$m x 5.2 $\mu$m. TEEM photonics waveguide mode field diameters were calibrated thanks to a reference fiber having a known $4.5$ $\pm$ $0.5$ $\mu$m mode field diameter.}
\end{figure}

%FIRST team collaborate with Grenoble Institute for Planetology and Astrophysics (IPAG) and TEEM Photonics based in Grenoble, France to develop the PIC. 
The FIRST PIC is developed in collaboration with TEEM Photonics based in Grenoble, France. TEEM photonics IonExt technology consists in $K_+:Na_+$ ion exchange in glass. A lithographic mask is put on the substrate surface and the ion exchange is realized into an electrolyzed bath. The straight waveguide propagation loss is about $0.3$ $dB/cm$ at $780$ $nm$.  The monomodal range is reached for wavelengths between $530$ and $820$ $nm$ for $2$ $\mu m$ wide waveguides, see Fig.~\ref{fig:cutoff}. The SM-630 fiber coupling loss is specified by Teem Photonics to be $0.5$ $dB$ for a $2$ $\mu m$ wide waveguide over the monomodal spectral band. 

Polarized mode field diameter measurements were performed. With a x40 objective, the waveguide mode is imaged on a camera and the intensity profile for an horizontal and vertical cut of the mode are presented in Fig.~\ref{fig:cutoff} and Table.~\ref{tab:MFD}. Polarized mode field diameter measurements show that the waveguide transmission is better in TE than TM. There is an asymmetry in the vertical profile, explained by the fact that diffusion is performed at the surface of the substrate and that waveguides are not buried into the glass, inducing a form birefringence.

 \begin{table}[ht]
\caption{Polarized mode field diameters measurements at $635$ $nm$. Mode width corresponds to the $1/e^2$ width of the mode horizontal profile shown in Fig.~\ref{fig:cutoff}. Because the mode is asymmetric in the vertical direction, two measurements are performed in this direction: the mode upper (resp. lower) height correspond to the upper (resp.lower) $1/e^2$ half-width of the mode in this direction.} 
\label{tab:MFD}
\begin{center}       
\begin{tabular}{|l|l|l|l|}
\hline
\rule[-1ex]{0pt}{3.5ex} Polarization & Mode width & Mode upper height & Mode lower height \\
\hline
\rule[-1ex]{0pt}{3.5ex} TE & 4.81 $\pm$ 0.64 um & 0.74 $\pm$ 0.17 um & 1.12 $\pm$ 0.42 um \\
\hline
\rule[-1ex]{0pt}{3.5ex} TM & 4.68 $\pm$ 0.61 um & 0.65 $\pm$ 0.15 um & 1.07 $\pm$ 0.29 um\\
\hline
\rule[-1ex]{0pt}{3.5ex} Mean & 4.75 $\pm$ 0.69 um & 0.70 $\pm$ 0.21 um & 1.10 $\pm$ 0.38 um\\
\hline
\end{tabular}
\end{center}
\end{table}

Kevin Barjot\cite{barjot2021} characterized the current 5T chip prototype: its throughput is about 15$\%$ to 30$\% $ depending on the type of combiner used. This low throughput is mainly due to non-optimized Y splitters and too small bend curvature radii.

\subsubsection{Modeling of the waveguides diffused index profile on Beamprop software}
\begin{figure} [h!]
   \begin{center}
   \begin{tabular}{c} %% tabular useful for creating an array of images 
   \includegraphics[height=3cm]{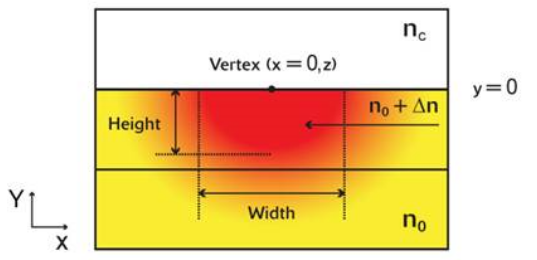}
   \end{tabular}
   \end{center}
   \caption[BeampropModel] 
%>>>> use \label inside caption to get Fig. number with \ref{}
   { \label{fig:Beampropmodel} Cross-section of the 3D built-in diffused index profile. \textit{Credit: Beamprop} }
   \end{figure}

Based on the cutoffs and mode field diameters measurements reported in Sec.~\ref{sec:Labmeas}, a 3D diffused index profile is derived to model the TEEM photonics waveguides with the Beamprop software. The model is presented in Fig.~\ref{fig:Beampropmodel}. The cross section of the diffused index profile $n(x,y)$ is defined by: 

\begin{equation}
\label{eq:fov}
n(x,y)=n_o+[\Delta ng(x)f(y)]^\gamma 
\end{equation}

with 

\begin{equation}
\label{eq:fov}
g(x)=\frac{1}{2}\left\{erf[(\frac{w}{2}-x)/h_x)]+erf[(\frac{w}{2}+x)/h_x]  \right\} 
\end{equation}

\begin{equation}
\label{eq:fov}
f(y)=exp(\frac{-y^2}{h_y^2})
\end{equation}

where $n_0=1.52$ at $635$ $nm$ is the background material refractive index and $n_c=1.49$ at $635$ $nm$ is the glue layer refractive index. The glue layer is about 20 $\mu m$ wide and can be considered as infinite in the vertical direction. $h_x$ and $h_y$ are the diffusion lengths in the horizontal and vertical direction respectively. 
%For this first model, the diffusion is assumed to follow the same characteristics in the x and y direction i.e $h_x=h_x$. 
For the sake of simplicity, the diffusion process is assumed to be equivalent in both x and y directions, i.e. $h_x=h_y$. Coefficient $\gamma$ represents the non-linear interaction between the ion concentration and the index. In this model $\gamma=1$. The parameters optimized thanks to the laboratory measurements are the waveguide width and height, $w$ and $h$, and the index difference $\Delta n$ between the substrate and the waveguide core. 

\subsubsection{Polarization behavior}

This section presents some preliminary results on TEEM photonics waveguide behavior in terms of polarization. 

\begin{wrapfigure}{r}{9cm}
%\begin{figure}
  \centering
  \includegraphics[height=7cm]{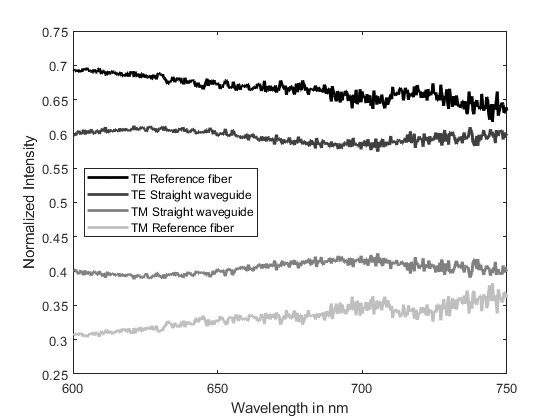}
\caption[RefStraight] 
{ \label{fig:RefStraight} Reference fiber and straight waveguide polarized intensity as a function of wavelength, normalized by total intensity (TE $+$ TM)}
%\end{figure}
\end{wrapfigure}

In Fig.~\ref{fig:RefStraight}, a P2-830A reference fiber is connected to a white SLED source to inject light into the PIC. A x4 objective is used so that the light at the PIC output feeds an OceanOptics spectrometer. A linear polarizer selects the TE (horizontal, in the plane of the PIC) or TM (vertical) polarization. All spectra were corrected from the spectrograph polarized diffraction efficiency.  
%integration time variation 
%This plot shows the reference fiber and straight waveguide polarized intensity as a function of wavelength, normalized by total intensity (TE $+$ TM). 
Fig.~\ref{fig:RefStraight} shows a 70:30 TE:TM ratio in the source. 
%passing through the reference fiber
After propagation into the straight waveguide, the TE:TM ratio is 60:40. This difference can be due to a misalignment between the polarizer and waveguide axis, or, between the polarizer axis from one measurement (the reference fiber one) to the other (the PIC one). These hypotheses are unlikely because the axes misalignment should be significant, i.e. around 18$^\circ$ ($cos^2(18)=0.90$). A leakage from TE to TM polarization is considered to be unlikely in a straight waveguide. Finally, the more reasonable explanation is that TE polarization might be more attenuated than TM polarization when propagating in the PIC. This result is consistent with Fig.~\ref{fig:cutoff}.

For now, the precision on mode field diameters measurements is not sufficient to run polarimetric Beamprop simulations. Effective indexes and throughput in both polarization will be further investigated thanks to the new chip characterization, see Sec.~\ref{sec:newWafer}.

\section{Laboratory characterization and optimization of FIRST PIC combiners}
\label{sec:FIRST_PICSCombiners}
\subsection{Directional couplers}

\subsubsection{Theoretical model}
\label{sec:theoryCoupler}
\begin{figure} [h]
\begin{center}
\begin{tabular}{l}
\includegraphics[height=6cm]{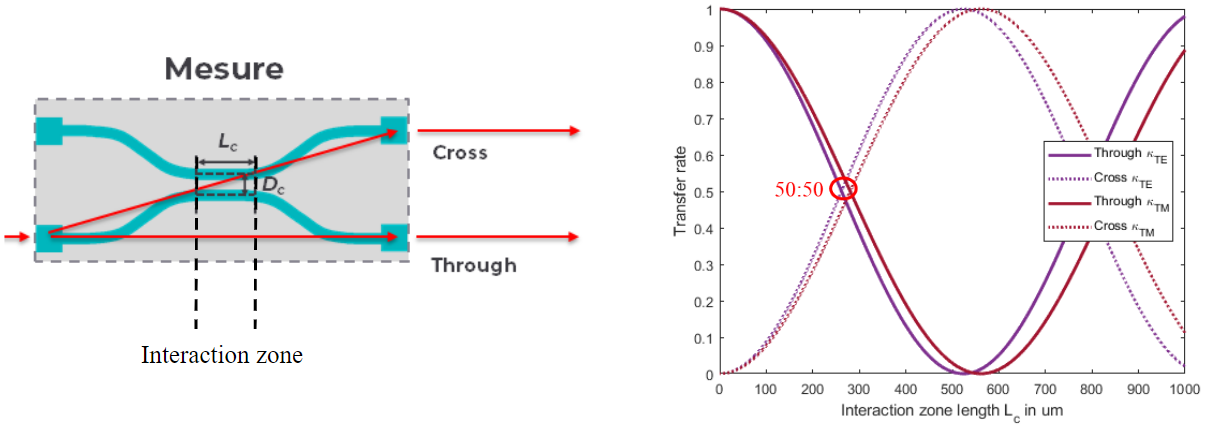}
\end{tabular}
\end{center}
\caption[X] 
%>>>> use \label inside caption to get Fig. number with \ref{}
{ \label{fig:Model} \textbf{Left:} Standard geometry of a directional coupler \emph{(credit: TEEM photonics)}. The interaction zone is defined by its gap value between the coupling waveguides $D_c$ and its interaction zone length $L_c$. \textbf{Right:} Theoretical transfer rate as a function of the interaction zone length $L_c$ for a uniformly symmetric directional coupler. The coupling coefficient depends on the polarization for birefringent waveguides.}
\end{figure}

The standard geometry of a directional coupler is presented in Fig.~\ref{fig:Model}. The output power in each arm of a directional coupler\cite{labeyephd} is given by: 
\begin{equation}
\label{eq:fov}
P1 = 1-(\kappa^2/\Delta^2)*sin^2(\Delta*L)
\end{equation}

\begin{equation}
\label{eq:fov}
P2 = (\kappa^2/\Delta^2)*sin^2(\Delta*L)
\end{equation}

with $\kappa$ the mode coupling coefficient and $\Delta$ defined as:  

\begin{equation}
\label{eq:fov}
\Delta=\sqrt{(\frac{\beta_1-\beta_2}{2})^2+\kappa^2} \quad \propto \quad \frac{1}{L_{max}}
\end{equation}
with $\beta_1$ and $\beta_2$ the propagation constants of the waveguides 1 ("Through" in Fig.~\ref{fig:Model}) and 2 ("Cross"), $L$ the interaction region length. $F=\kappa^2/\Delta^2$ is the maximum power transferred from a waveguide to the other. This maximum power transfer is performed for an interaction zone length $L_{max}$. 

The directional coupler transfer rate is defined as the proportion of the output flux in each output. The specification for the PIC combiners is to get a transfer rate of 50:50 (50$\pm$10) over the 600 to 800 $nm$ spectral band. 
For a uniformly symmetric directional coupler, i.e waveguide 1 and 2 are identical in the coupling zone, $\beta_1=\beta_2$ so $F=1$, meaning that all the power in one waveguide can be transferred into the other. To get a 50:50 transfer rate, the interaction region length $L_{50:50}$ is given by: 

\begin{equation}
\label{eq:fov}
\frac{\Delta L_{50:50}}{\Delta L_{max}}= \frac{(2k+1)\frac{\pi}{4}}{\frac{\pi}{2}}\quad \textrm{so} \quad L_{50:50}=(k+\frac{1}{2})L_{max} \quad \textrm{with} \quad k\in \mathbb{N}
\end{equation}

As the coupling length $L_{max}$ depends on the wavelength, the transfer rate of a uniformly symmetric coupler is chromatic\cite{labeyephd}. For asymmetric directional couplers,  $\beta_1(\lambda)\neq\beta_2(\lambda)$ and one can compensate the directional coupler chromaticity.

\subsubsection{Symmetric couplers: laboratory characterization}
\begin{figure} [h]
\begin{center}
\begin{tabular}{ll}
\includegraphics[height=15.5cm]{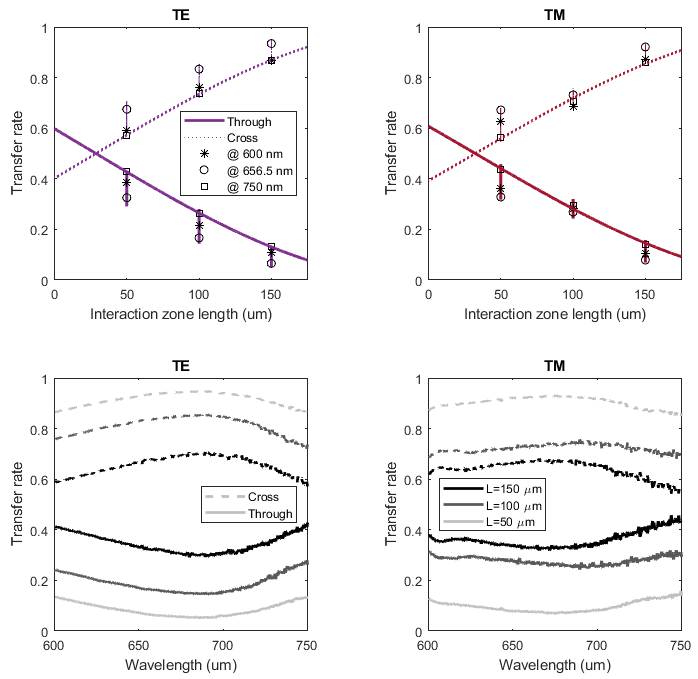}
\end{tabular}
\end{center}
\caption[X] 
%>>>> use \label inside caption to get Fig. number with \ref{}
{ \label{fig:TransferTETM} \textbf{Top:} Laboratory transfer rate measurement as a function of the interaction zone length. The measurement is performed using a P2-830A reference fiber connected to a white SLED source to inject light into the PIC. A x4 objective is used so that the light at the PIC output feeds an OceanOptics spectrometer. A linear polarizer selects the TE (horizontal, in the plane of the PIC) or TM (vertical) polarization. Vertical lines show the maximum variation of the transfer rate over the whole spectral band and some particular wavelengths are highlighted with markers. The curves correspond to the fit with the theoretical model presented Sec.~\ref{sec:theoryCoupler}. \textbf{Bottom:} Laboratory transfer rate measurement as a function of wavelength for various interaction zone lengths. These symmetric directional couplers are not highly chromatic, in particular for the TM polarization in the $600$-$700$ $nm$ range.}
\end{figure}

Three uniformly symmetric directional couplers have been fabricated. Every coupler has a $D_c=2$ $\mu m$ gap but a different interaction zone length: $L_c=50$ $\mu m$, $100$ $\mu m$ or $150$ $\mu m$. These uniformly symmetric directional couplers have been characterized in terms of transfer rate in TE and TM polarization and transfer rate chromaticity, and the results are reported in Fig.~\ref{fig:TransferTETM}. The objectives are 1) find the interaction zone length $L_{50:50}$ to get a 50:50 (50$\pm$10) transfer rate and 2) investigate the transfer rate chromaticity over our spectral band. 

Measurements in Fig.~\ref{fig:TransferTETM} confirmed that the coupling coefficient depends on polarization, i.e waveguides are birefringent. It is not an issue for our application as long as there is no cross-talk between polarization leading to fringes blurring. TE polarization could be more attenuated than TM polarization into the PIC too, see Sec.~\ref{sec:newWafer} for discussion. Fig.~\ref{fig:TransferTETM} also shows that a compromise can be found for the interaction zone length which is situated between 20 and 40 um depending on polarization for the symmetric directional coupler. New symmetric directional couplers are fabricated to probe this interaction zone length range.

\subsubsection{Asymmetric couplers: Beamprop simulations}
\begin{figure} [h]
\begin{center}
\begin{tabular}{ll}
\includegraphics[height=9.5cm]{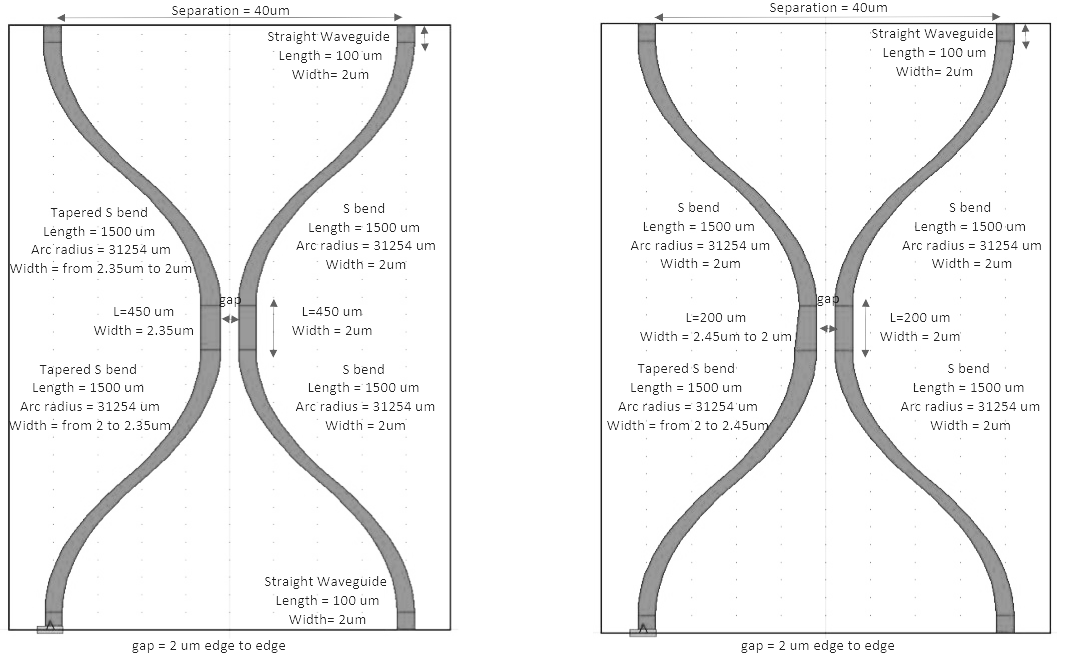}
\end{tabular}
\end{center}
\caption[CouplerAsym] 
%>>>> use \label inside caption to get Fig. number with \ref{}
{ \label{fig:XasymScheme} Asymmetric directional coupler geometries optimized with the Beamprop module of Rsoft software.\textbf{ Left:} Uniformly asymmetric directional coupler.\textbf{ Right:} Non symmetric directional coupler. The transfer rate specfication is 50:50 with an accuracy of $\pm 10$ over the 600 to 850nm spectral band.}
\end{figure}

\begin{figure} [h]
\begin{center}
\begin{tabular}{ll}
\includegraphics[height=5cm]{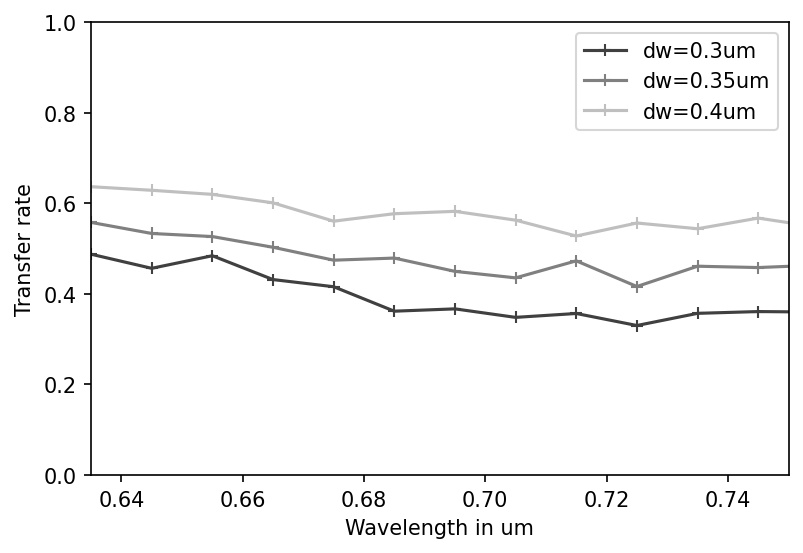}&\includegraphics[height=5cm]{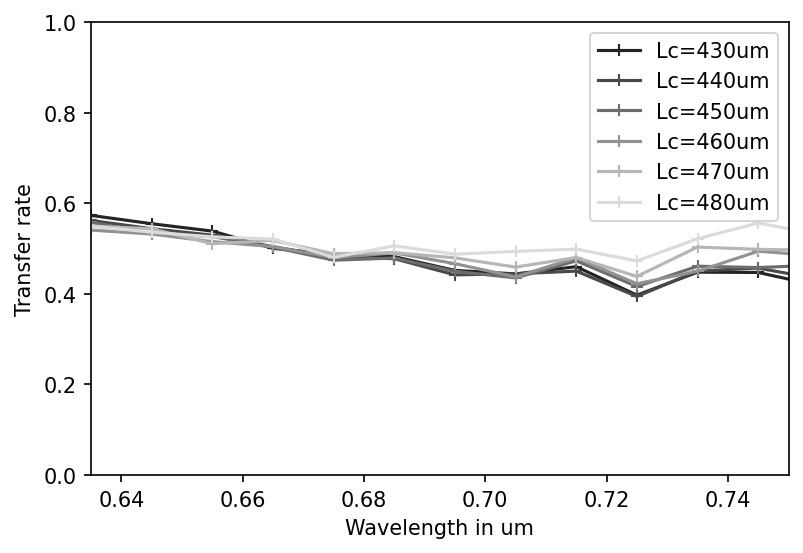} 
\end{tabular}
\end{center}
\caption[CouplerAsym] 
%>>>> use \label inside caption to get Fig. number with \ref{}
{ \label{fig:SimulationsXUA} Uniformly asymmetric directional coupler (see Fig.~\ref{fig:XasymScheme}) simulated transfer rate. For the sake of readability, the fraction of flux is plotted for only one of the directional coupler arms. \textbf{Left:} transfer rate as a function of waveguides differential width in the interaction region of length $L_c=450$ $\mu m$. \textbf{Right:} transfer rate as a function of the interaction region length $L_c$ for waveguides differential width $dw=0.35$ $\mu m$.}
\end{figure}

\begin{figure} [h]
\begin{center}
\begin{tabular}{ll}
\includegraphics[height=5cm]{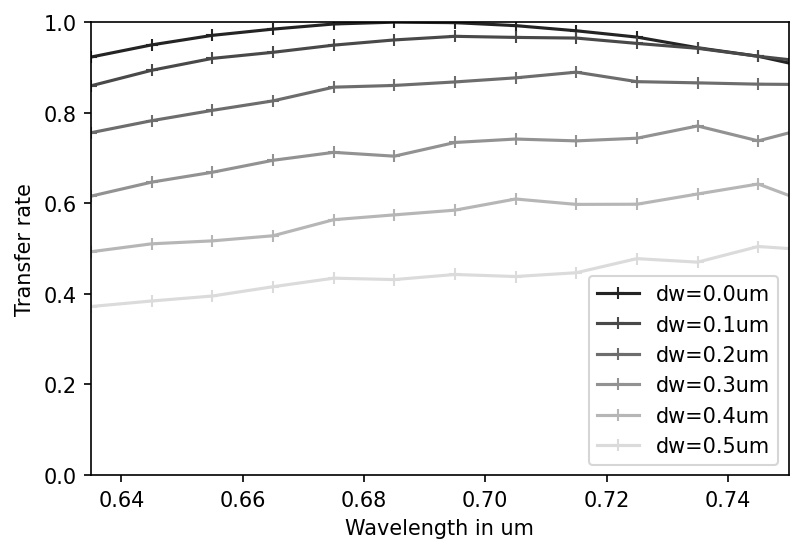}&\includegraphics[height=5cm]{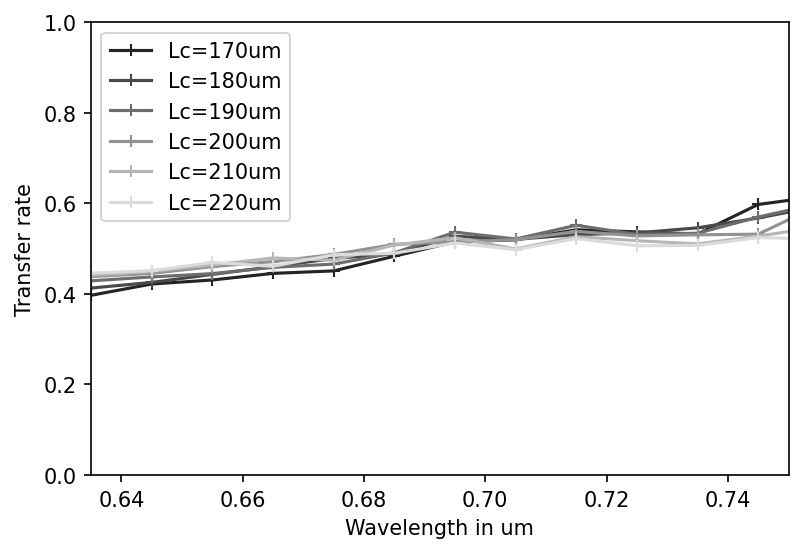} 
\end{tabular}
\end{center}
\caption[CouplerAsym] 
%>>>> use \label inside caption to get Fig. number with \ref{}
{ \label{fig:SimulationsXNS} Non symmetric directional coupler (see Fig.~\ref{fig:XasymScheme}) simulated transfer rate. For the sake of readability, the fraction of flux is plotted for only one the directional coupler arms. \textbf{Left:} transfer rate as a function of waveguides differential width in the interaction region of length $L_c=200$ $\mu m$. \textbf{Right:} transfer rate as a function of the interaction region length $L_c$ for waveguides differential width $dw=0.45$ $\mu m$.}
\end{figure}

This section presents Beamprop simulations of the spectral transfer rate for two asymmetric couplers defined as uniformly asymmetric (UA) and non symmetric (NS)\cite{Takagi1992}.
The geometry of the simulated asymmetric couplers is shown in Fig.~\ref{fig:XasymScheme}. A uniformly asymmetric directional coupler is composed of one waveguide wider than the other by an amount $dw$ in the interaction zone. For non symmetric directional coupler, one waveguide is wider than the other in the interaction zone and is also tapered. The parameter space ($dw$: differential width between waveguides in the interaction zone, $L_c$: the interaction zone length) is probed to define optimal solutions for each geometry, see values in Fig.~\ref{fig:XasymScheme}. Then, the transfer rate chromaticity is simulated for various width difference and interaction zone length around the optimal parameter doublet ($dw$,$L_c$) for each geometry. This study is performed for tolerancing purposes, knowing that the lithographic precision on waveguide width is about $0.1$ $\mu m$. Fig.~\ref{fig:SimulationsXUA} shows that the transfer rate for UA couplers is more sensitive to width difference between the waveguides than interaction zone length. From Fig.\ref{fig:SimulationsXNS}, one can derive that these NS couplers are also more sensitive to fabrication errors on the width difference. With a directional coupler, fringes are sampled with two outputs, so one still needs to piston the segmented mirror to scan the fringes. A minimum of 2 images per baseline is therefore required. For a large amount of baselines, many images are thus needed, which takes time. A way to increase the acquisition efficiency is to use ABCDs.

\subsection{Design of an achromatic $\pi/2$ passive phase shifter for ABCD combination}
\subsubsection{Theoretical model}
\label{subsubsec:theoryABCD}
The ABCD beam combination is an interferometric scheme\cite{labeyephd,Benisty2009} allowing fringe sampling at its output. An ABCD is composed of two inputs and 4 outputs each being the recombination of the two inputs with a different phase as represented in Fig.~\ref{fig:ABCD}. Both inputs, i.e interfering beams, are split in two and are recombined by pairs thanks to two directional couplers. The phases at the ABCD outputs are specified as follows: 

\begin{figure} [h]
\begin{center}
\begin{tabular}{c} %% tabular useful for creating an array of images 
\includegraphics[height=3cm]{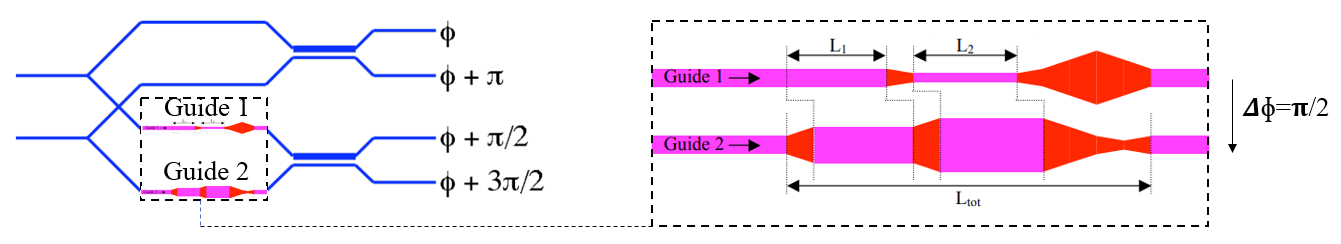}
\end{tabular}
\end{center}
\caption[ABCD] 
%>>>> use \label inside caption to get Fig. number with \ref{}
{ \label{fig:ABCD} ABCD recombination scheme adapted from Pierre Labeye PhD thesis and Benisty et al. 2009\cite{labeyephd,Benisty2009}.}
\end{figure}

\begin{equation}
\begin{aligned}
\label{eq:ABCD}
& A: \varphi_{12}^{A} \\
& B: \varphi_{12}^{B}=\varphi_{12}^{A}+\pi \\
& C: \varphi_{12}^{C}=\varphi_{12}^{A}+\pi/2\\
& D: \varphi_{12}^{D}=\varphi_{12}^{A}+\pi/2+\pi=\varphi_{12}^{A}+3\pi/2
\end{aligned}
\end{equation}

In order to build an achromatic ABCD, a $\pi/2$ achromatic passive phase shifter is required. A phase shifter is composed of two parallel waveguides: arm 1 and 2. Each arm is divided into N segments of optimized widths and lengths. Both arms have the same total length. The optical path is modified from one arm to the other thanks to the width variations, which induce variation of the mode effective index. The phase difference for a phase shifter of length L composed of N=1 segment expresses as: 
% effective refractive index

\begin{equation}
\label{eq:phaseshifter}
\Delta\varphi=\varphi_1-\varphi_2=\frac{2\pi}{\lambda}(n_{eff,1}-n_{eff,2})L
\end{equation}
with $n_{eff,1}=A_1+B_1\lambda+C_1\lambda^2$ and $n_{eff,2}=A_2+B_2\lambda+C_2\lambda^2$ the effective indexes of waveguides 1 and 2. The phase shifter is achromatic if: 
\begin{equation}
\label{eq:fov}
\forall \lambda, \frac{\partial \varphi}{\partial \lambda}=-\frac{2\pi}{\lambda}(\frac{\Delta n}{\lambda}-\frac{\partial \Delta n}{\partial \lambda})L=0 
\end{equation}

Following P. Labeye PhD thesis\cite{labeyephd}, these equations can be generalized for a design comprising N segments shifting the phase by $\varphi_0=\pi/2$:

%Equation.~\ref{eq:NphaseshifterEq1} and Equation.~\ref{eq:NphaseshifterEq2} describe a N segments $\varphi_0=\pi/2$ phase shifter achromatic on the $600$ to $800$ $nm$ spectral band.\cite{labeyephd}

\begin{equation}
\label{eq:NphaseshifterEq1}
\varphi=\frac{2\pi}{\lambda}\sum_{i=1}^{N}\Delta n_i L_i=\varphi_0
\end{equation}

\begin{equation}
\label{eq:NphaseshifterEq2}
\frac{\partial \varphi}{\partial \lambda}=-\frac{2\pi}{\lambda}\sum_{i=1}^{N}(\frac{\overline{A_i}}{\lambda}-\overline{C_i}\lambda)L_i=0
\end{equation}
with $\Delta n_i=n_{eff,i1}-n_{eff,i2}=(A_{wi1}-A_{wi2})+(B_{wi1}-B_{wi2})\lambda+(C_{wi1}-C_{wi2})\lambda^2=\overline{A_i}+\overline{B_i}\lambda+\overline{C_i}\lambda^2$.

Equation~\ref{eq:NphaseshifterEq2} can be true for all wavelengths only if:
%$\sum_{i=1}^{N}(\frac{\overline{A_i}}{\lambda}-\overline{C_i}\lambda)L_i=0$ is true for all wavelengths if: 

\begin{equation}
\label{eq:NphaseshifterEq3}
\sum_{i=1}^{N}\overline{C_i}L_i=0 \quad \textrm{and} \quad \sum_{i=1}^{N}\overline{A_i}\L_i=0.
\end{equation}

As a consequence, one can derive that:

\begin{equation}
\label{eq:NphaseshifterEq4}
\sum_{i=1}^{N}(\overline{B_i}L_i)=\frac{\varphi_0}{2\pi}
\end{equation}

is also a condition for Eq.~\ref{eq:NphaseshifterEq1} to be valid for all wavelengths.

It is thus possible to find sets of $(\overline{A_i}, \overline{B_i}, \overline{C_i})$ parameters that will produce an achromatic phase shifter over a given bandwidth. 

%For a given waveguide width, the coefficients $A_i$, $B_i$ and $C_i$ are estimated based on simulations performed with the Beamprop software. By optimizing the segment widths and lengths, it is possible to optimize the phase shifter chromaticity over a given bandwidth.

%\subsubsection{Beamprop simulation}
\subsubsection{Optimization process}

\begin{figure} [h!]
\begin{center}
\begin{tabular}{c} %% tabular useful for creating an array of images 
\includegraphics[height=5cm]{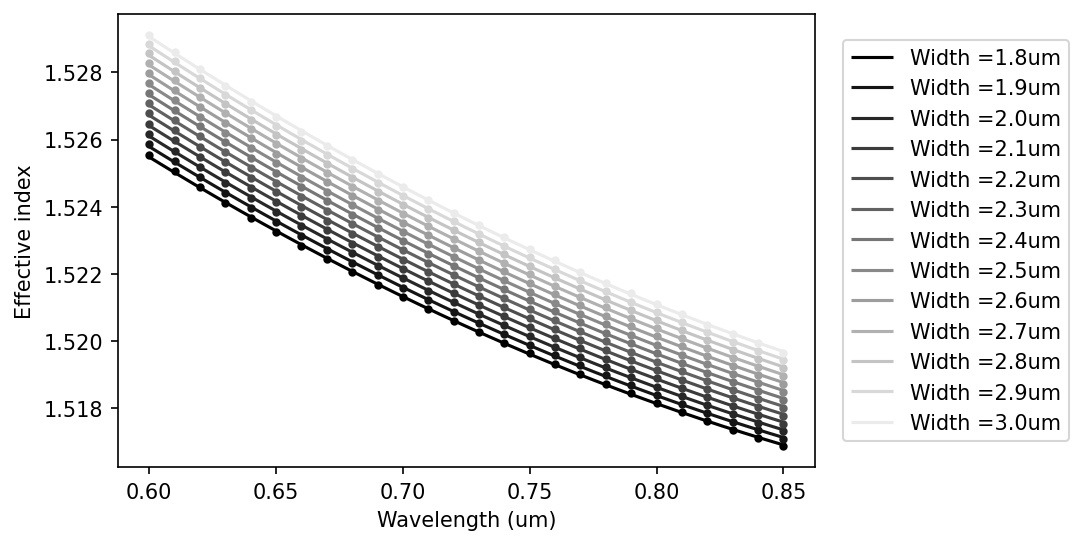}
\end{tabular}
\end{center}
\caption[BeampropNeff] 
%>>>> use \label inside caption to get Fig. number with \ref{}
{ \label{fig:BeampropNeff} Beamprop computation (dots) and fit (lines) of the waveguide effective index as a function of wavelength and waveguide width}
\end{figure}

For a given waveguide width, the coefficients $A_i$, $B_i$ and $C_i$ are estimated based on simulations performed with the Beamprop software. %By optimizing the segment widths and lengths, it is possible to optimize the phase shifter chromaticity over a given bandwidth.
Effective indexes are computed for widths between $1.8$ and $3$ $\mu m$ with a step of $0.1$ $\mu m$, as shown in Fig.~\ref{fig:BeampropNeff}. A second order polynomial fit is used to estimate the $A_i$, $B_i$ and $C_i$ coefficients for each waveguide width value. Two- and three-segments solutions for achromatic $\varphi_0=\pi/2$ phase shifters are investigated. 

The system of equations to be solved is given by Eq.~\ref{eq:NphaseshifterEq3} and Eq.~\ref{eq:NphaseshifterEq4}, which can be rewritten with a matrix formalism: 
\begin{equation}
\label{eq:System}
\left[ L \right]=\left[ M \right]^{-1}\cdot \left[ \phi \right]
\end{equation}

with $M=\begin{bmatrix}
\overline{A_1} & \overline{A_2} & \overline{A_3}\\
\overline{B_1} & \overline{B_2} & \overline{B_3}\\
\overline{C_1} & \overline{C_2} & \overline{C_3}\\
\end{bmatrix}$ and $\left[ \phi \right]=\begin{bmatrix}
0\\
\frac{\varphi_0}{2\pi}\\
0\\
\end{bmatrix}$ for the three-segment solution. 

with $ M=\begin{bmatrix}
\overline{A_1} & \overline{A_2}\\
\overline{B_1} & \overline{B_2}\\
\end{bmatrix}$ and $\left[ \phi \right]=\begin{bmatrix}
0\\
\frac{\varphi_0}{2\pi}\\
\end{bmatrix}$ for the two-segment solution. 

%To solve the systems, effective indexes are computed on Beamprop for widths between $1.8$ and $3$ $\mu m$ with a step of $0.1$ $\mu m$ shown Fig.~\ref{fig:BeampropNeff}. A second order polynomial fit is used to recover $A$, $B$ and $C$ coefficients for each waveguide width value. 

%\begin{figure} [h!]
%\begin{center}
%\begin{tabular}{c} %% tabular useful for creating an array of images 
%\includegraphics[height=6cm]{index.jpg}
%\end{tabular}
%\end{center}
%\caption[BeampropNeff] 
%>>>> use \label inside caption to get Fig. number with \ref{}
%{ \label{fig:BeampropNeff} Beamprop computation (dots) and fit (lines) of the waveguide effective index as a function of wavelength and waveguide width}
%\end{figure}

For the N=3 and N=2 segments phase shifters, one must pick six and four waveguide widths values respectively. This is done through a python code computing segments lengths for all the possible widths combinations. The shorter achromatic phase shifters solutions are kept. Waveguides are monomodal from $600$ to $820$ $nm$ for widths between $1.8$ and $2.3$ $\mu m$. Resolving Equation.~\ref{eq:System} leads to achromatic $\varphi_0=\pi/2$ phase shifters about a dozen millimeters long. That is why multi-modal segments are used: effective index increases with the waveguide width. For higher widths values, the light propagation is slowed down and the $\pi/2$ phase shift between the phase shifter's arm is reached for shorter total length. Mode coupling between segments is performed with tapers. Each taper present in one waveguide must be added to the other for phase matching\cite{labeyephd}. Simulations are run in the three following cases: without tapers, 10\,$\mu m$ long and 100 $\mu m$ long tapers.

\begin{figure} [h!]
\begin{center}
\begin{tabular}{ll}
\includegraphics[height=5.5cm]{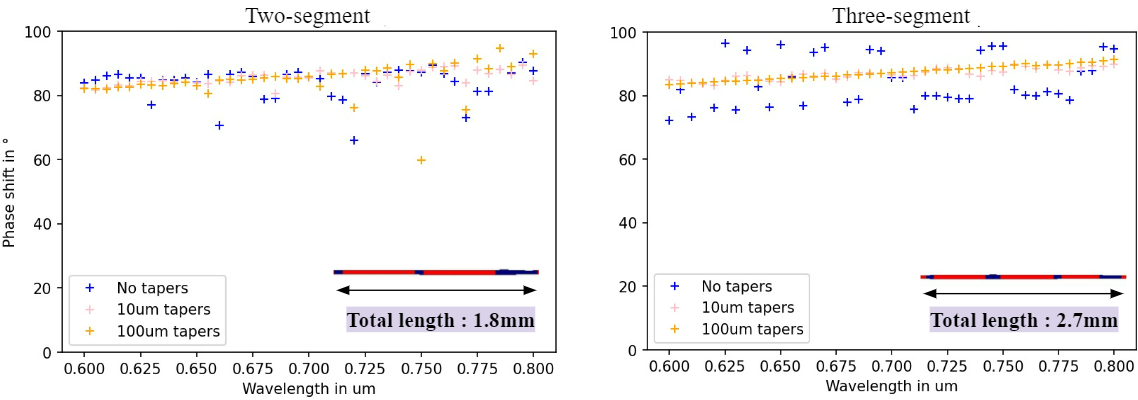}
\end{tabular}
\end{center}
\caption[PhaseShifters] 
%>>>> use \label inside caption to get Fig. number with \ref{}
{ \label{fig:Solutions} Phase difference between the $\pi/2$ phase shifter arms as a function of wavelength. \textbf{Left:} Two-segment solution. \textbf{Right:} Three-segment solution.}
\end{figure}

Results are reported in Fig.~\ref{fig:Solutions}, showing that an achromatic phase shift is achieved over the 600-800nm bandwidth and that tapers help to balance the phase shift.  The three-segment solution offers a better phase performance at the cost of a longer total length.

\vspace{3cm}

\section{Prospects}
\label{sec:newWafer}

\begin{figure} [h!]
\begin{center}
\begin{tabular}{l}
\includegraphics[height=11cm]{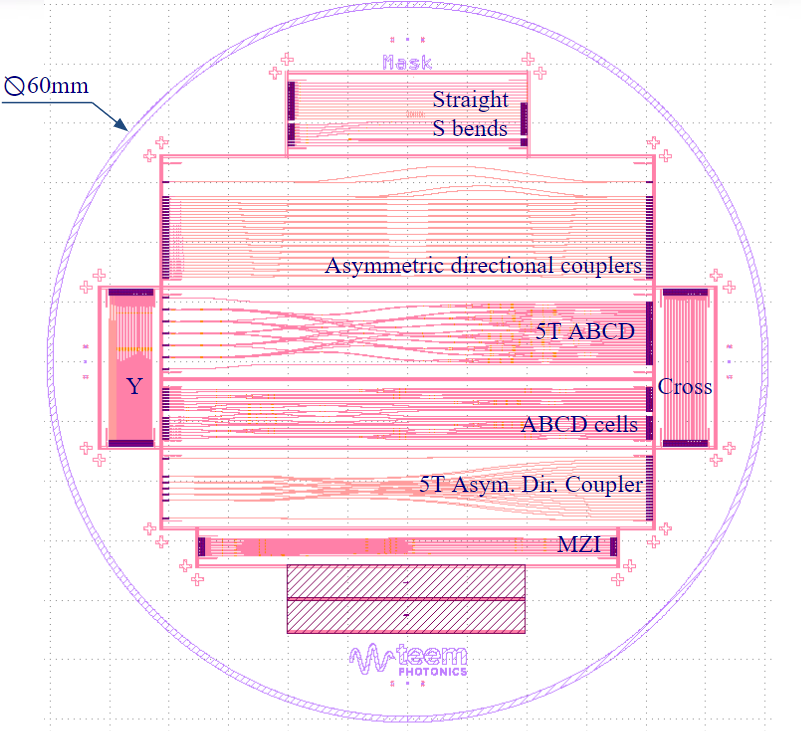}
\end{tabular}
\end{center}
\caption[Wafer] 
%>>>> use \label inside caption to get Fig. number with \ref{}
{ \label{fig:newwafer} New wafer GDS view on Klayout software. "Straight": straight waveguides with various widths. "S bends": S bend with various curvature radii. "MZI": Mach-Zehnder interferometers. "Cross": straight waveguides with various numbers of crossings at various angles. "Y": Y splitters with different splitting zone geometries.}
\end{figure}

Following this work, a new wafer presented in Fig.~\ref{fig:newwafer} has been designed and will be ready to be characterized by the end of August 2022. 
The objective is to test the designs presented in this proceeding. Two 5T PICs for FIRST interferometric combination are added to the wafer: one with asymmetric directional couplers and one with ABCD combiners. Concerning the previous designs, phase chromaticity of the asymmetric directional couplers and mode coupling between multi-modal and monomodal segments in ABCDs will be further investigated. Y junctions, S bends and crossings will be characterized. Some fundamental values of TEEM technology are not well known for now, especially concerning birefringence. Polarized effective index measurements will be performed thanks to a Mach-Zehnder interferometers (MZI). Form birefringence, increasing as a function of mask opening width, will be further studied with straight waveguides of various widths. This will be investigated especially for achromatic directional couplers and achromatic phase shifters in which waveguides widths are varying. 

% -straight waveguides with various widths = form birefringence will be investigated (increases as a function of mask opening width) = important to know as waveguides widths are changing in achromatic directional couplers and achromatic phase shifters. Polarized mode field diameter measurements (shows birefringence in straight waveguides), bi directional couplers polarized behavior (not the same coupling for a 50/50 transfer rate = birefringence). Does the birefringence changes depending on the building block ? results polarization cross talk at high bendings. Hypothesis to be tested (stress and form birefringence)\\ Currently: polarization issues in 5T with X directional couplers combiners (two chips with one polarization in each chip ?). 5T with Y junction combiners better but half of the interferometric signal is lost. Mode coupling between multi-modal and monomodal segments is to be further investigated. The waveguide height is assumed to be constant at 2um. That is a point to be verified too. Polarization behavior to be understood; recursive process \\

\section{Conclusion}
\label{sec:Ccl}
In this work, we presented the FIRST instrument, its new scientific objective and TEEM Photonics $K_+:Na_+$ ion exchange technology. TEEM Photonics waveguides are characterized in terms of cutoff wavelengths and polarized mode field diameters leading to the modeling of the waveguide diffused index profile on the Beamprop software. In order to meet FIRST beam combiner specifications: 1) symmetric directional couplers transfer rates are measured, 2) asymmetric directional couplers and ABCD combiners are simulated and optimized using the Beamprop software. A new wafer is currently being fabricated. Building blocks will be further investigated through a recursive process composed of simulations and laboratory characterizations.\\

\acknowledgments % equivalent to \section*{ACKNOWLEDGMENTS}       
I would like to thank TEEM Photonics for their support and trust. This project is supported by the French National Research Agency (ANR-21-CE31-0005) and the doctoral school Astronomy $\&$ Astrophysics of Ile de France (ED 127). 
% References
\bibliography{report} % bibliography data in report.bib
\bibliographystyle{spiebib} % makes bibtex use spiebib.bst
\end{document}